# A Reliable User Authentication and Data Protection Model in Cloud Computing Environments

Mohammad Ahmadi[1], Mostafa Vali[1], Farez Moghaddam[1], Aida Hakemi[2], Kasra Madadipouya[1]
[1]Faculty of Computing, Asia Pacific University of Technology and Innovation (APU), Malaysia
[2]Faculty of Computing, Universiti Teknologi Malaysia, Malaysia
dr.ahmadi@apu.edu.my, mostafa.hajv@yahoo.com.my, info@nerocloud.org, hakemiaida@gmail.com, kasra_mp@live.com

*Abstract*— Security issues are the most challenging problems in cloud computing environments as an emerging technology. Regarding to this importance, an efficient and reliable user authentication and data protection model has been presented in this paper to increase the rate of reliability cloud-based environments. Accordingly, two encryption procedures have been established in an independent middleware (Agent) to perform the process of user authentication, access control, and data protection in cloud servers. AES has been used as a symmetric cryptography algorithm in cloud servers and RSA has been used as an asymmetric cryptography algorithm in Agent servers. The theoretical evaluation of the proposed model shows that the ability of resistance in face with possible attacks and unpredictable events has been enhanced considerably in comparison with similar models because of using dual encryption and an independent middleware during user authentication and data protection procedures.

## I. INTRODUCTION

Cloud computing is an emerging service that use the benefits of modern technologies (e.g. grid computing, clustering, virtualization, and processing power) to store and share resources via pool of resources. Cloud computing services have considerable benefits that enhance the efficiency and reliability of on-demand IT services. However, numerous challenging issues face cloud computing and have attracted the attention of many researchers and service providers [1].

Data management, resource allocation, security, privacy and access controls, load balancing, scalability, availability and interoperability are the most challenging issues in cloud-based environments that have affected the reliability of the newfound technology [2]. These concerns have been classified to various parts and the most important part is ensuring about the user authentication processes [3] and managing authorized and un-authorized accesses when users outsource sensitive data share on public or private cloud servers [4].

There are two main processes that investigate the procedure of secure and reliable user authentication in cloud-based environments:

- Investigating unique identifiers of users during the initial registration phase.
- User authentication and validating user legal identities and acquiring their access control privileges for the cloud-based resources and services during the service operation phase [5].

These two procedures have been faced with several challenges regarding to security issues and scalability concerns according to the nature of cloud computing environments. Hence, an efficient user authentication model has been presented in this paper to enhance the rate of security and reliability in cloud-based services.

## II. RELATED WORKS

According to the importance of user authentication functionality in cloud computing environments, several models and algorithms have been presented in recent years.

An efficient user authentication framework was suggested at 2011 [6]. The main aim of that model was ensuring about the verification of user legitimacy before enter into a cloud environment by providing identity management, mutual authentication, session key establishment between the users and cloud server. The presented scheme could resist many popular attacks such as replay attack, man in the middle attack, and denial of service attack. However, the computational costs and scalability of the model were affected by this resistance.

In 2013, a user authentication scheme on multi-server environments for cloud computing were presented by Yang *et al.* [7]. The suggested framework could be applied to multi-server environments because the ID-based concept that was used. Hence, the rate of efficiency, security, and flexibility in user authentication procedure were improved and the computational costs were decreased in comparison with similar models.

A dynamic ID-Based remote mutual authentication model [8] based on Elliptic Curve Cryptosystem (ECC) was proposed by Tein-Ho *et al.* in 2011. Subsequently, a Cloud Cognitive Authenticator (CCA) [9] was proposed in 2013 based on integrated authentication functionality. CCA uses the concepts of one round Zero Knowledge Proof (ZKP) and Advance Encryption Standard (AES) to enhance the security in public, private or hybrid clouds by four procedures providing with two levels of authentication and encrypting the user identifiers. The main specification of CCA in comparison with other models is the coverage of the two levels of authentication together with strength of the encryption algorithm. However, interoperability and compatibility with AES are the major weaknesses of CCA.

Yang and Lin [10] proposed an ID-based user authentication model by introducing three roles in the model: the user, the server, and the ID provider. The main responsibility of ID provider is to generate the registration and authentication information for both user and server. Moreover, two main phases have been presented regarding to the described investigation procedures: the registration phase and the mutual authentication phase. This model is compatible with various cloud environments and considerably cheaper in comparison with other models.

In 2013 an agent-based user authentication model in cloud computing environments was introduced [11] to increase the performance of user authentication processes according to the concept of agent. The theoretical analysis of this model shows that the suggested model increases the reliability and rate of trust in cloud-based environments. However, the idea of using agents in the process of authentication can be more efficient and reliable according to the capabilities of agents.

In 2014, Fatemi Moghaddam *et al.* presented a scalable user authentication model based on the concept of multi authentication and two main agents [5]: a client-based user authentication agent for confirming identity of the user in client-side, and a cloud-based software-as-a-service application for confirming the process of authentication for un-registered devices. The theoretical analysis of the suggested scheme showed that, designing this user authentication and access control model have enhanced the reliability and rate of trust in cloud computing environments. However, the computational costs were still a challenging issue in this model.

### III. Definitions

The proposed model has been suggested in this part by combination of two cryptography algorithms and other technologies for improving the security and reliability of user authentication procedures in cloud computing environments. Accordingly, following concepts should be defined:

*A. Agent*

The concept of agent in the proposed model is an independent middleware between the end-user and cloud servers to authorize user accesses and manage these procedures.

*B. Main Cryptography*

The main cryptography is a cloud-based symmetric encryption algorithm that is absolutely independent of end-user's performance. The main cryptography is based on AES.

*C. Secondary Cryptography*

An asymmetric cryptography algorithm that is completely based on users' performance to manage user authentication procedures and control accesses. The secondary cryptography is based on RSA.

### IV. Proposed Model

The proposed model has been designed to manage accesses and track the performance of data transmission between cloud servers and end users. Fig. 1 shows the suggested model in brief.

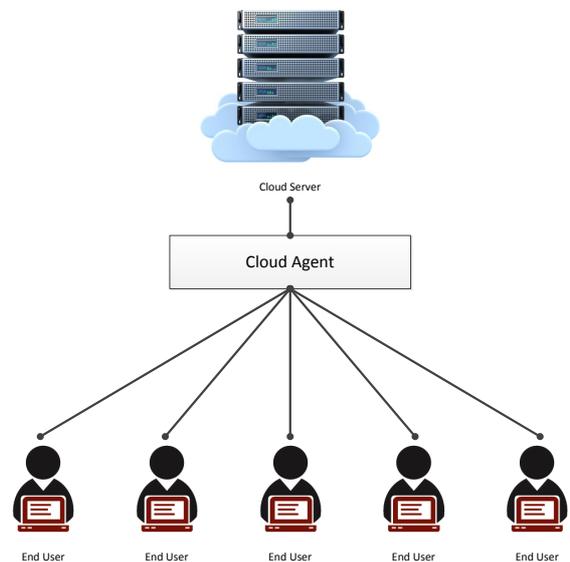

**Fig. 1.** The Proposed Model in Brief.

Regarding to the nature of data in cloud storages, data is classified to three main categories: Public, Private and Shared. As was described the performance of main cryptography is absolutely independent of the performance of end-users or the characteristics of data. The main cryptography procedure is done in cloud servers with AES-256. Regarding to the nature of main cryptography, a symmetric key encryption is most appropriate for this process.

The secondary cryptography establishes a secure connection between end-users and cloud servers for user authentication, data transmission, access controls. In fact, the key of main cryptography is re-encrypted by a RSA to protect the main cryptography procedure. Fig. 2 shows the performance of main and secondary cryptography procedures in brief.

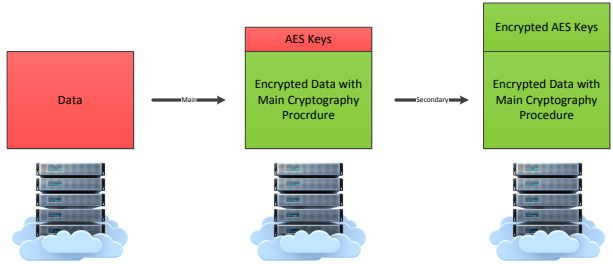

**Fig. 2.** Performance of Main and Secondary Cryptography

As was described, Agent is an independent middleware between the cloud servers and end users to establish secure connection between the cloud server and end-users. Agents in the proposed model have their own servers and cloud storages to define and store access rules and keys. There are four main responsibilities for Agent in the suggested model.

*A. Secondary Cryptography*

The process of secondary cryptography has been shown in Fig. 3. The RSA keys are generated in Agent and the public key is transferred to the cloud server. After encrypting the AES key with the RSA public key, the public and private key of RSA are stored in Agent servers and a copy of public key is transferred to data owner.

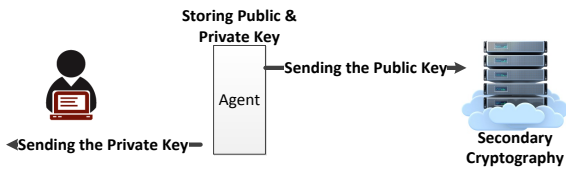

**Fig. 3.** Secondary Cryptography.

*B. User Authentication of Data Applicant*

According to Fig. 4, the request of data applicant is encrypted with the own private key and is sent to the Agent. The Agent decrypts the request with the public key and verifies the identity of data applicant.

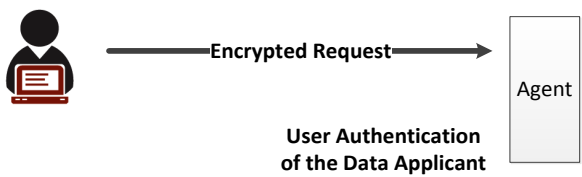

**Fig. 4.** User Authentication of the Data Applicant

*C. User Authentication of Data Owner*

The access request is sent to the data owner. Data owner verifies the request and encrypts the verification twice with his private key and the data private key and sends them to the Agent. Agent verifies the identity of data owner with decrypting data by the public key of data owner. Furthermore, the verification is submitted by decrypting data by the public key of data that was stored in Agent's cloud storage before. Fig. 5 Shows this process in details.

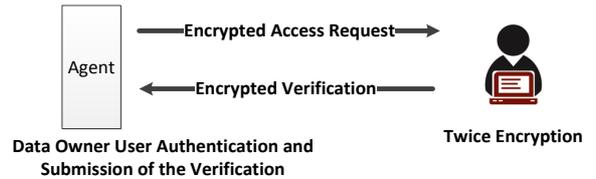

**Fig. 5.** Data Owner User Authentication

*D. Managing Accesses to Main Cloud Servers*

The private key of data is encrypted by the private key of the user and is sent to the cloud server. In cloud server, the private key of data is decrypted by the public key of user, the AES main key is decrypted by the private key of data, and the data is decrypted by the AES main key. Fig. 6 shows these procedures in details.

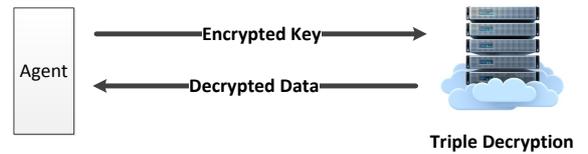

**Fig. 6.** Managing Accesses to Main Cloud Servers

V. DISCUSSION

*A. Security Justification*

The security justification of the suggested model has been evaluated according to the following table:

TABLE I: SECURITY JUSTIFICATION OF THE PROPOSED MODEL

| Challenges | Issues | Reasons |
|---|---|---|
| Data protection in servers | Losing Data | Un-Secure Cryptography |
| Secure Data Transmission | Losing Data | Un-Secure Transmission |
| User Authentication | Losing Data | Un-Secure Authentication |
| User Authentication | Losing Server | Un-Predictable Attacks |
| Access Controls | Un-Authorized | Un-Reliable Algorithm |
| Access Controls | Un-Authorized | Lack of Scalability |
| Lack of Resistance | Losing Data | Un-Efficient Resistance |

*B. Security Analysis*

The process of security analysis was considered as follows:

*1) Two-Step Cryptography*

By using two symmetric and asymmetric cryptography algorithms the reliability of the system is enhanced considerably. Accordingly, with a failure of one cryptography algorithm during unpredictable events or attacks the security of the system is guaranteed with the other cryptography algorithm and the time for resistance is provided. Furthermore, by using two step of cryptography for several security procedures (i.e. user authentication, data protection in cloud servers, data protection in data transmission, key exchange and key generation) the efficiency of this security model is enhanced significantly.

*2) Main Cryptography*

The powerful AES cryptography algorithm is responsible for the main cryptography procedure and because of the stability of the keys and the lack of transmission in all scenarios of this model; this symmetric algorithm is the most appropriate algorithm for main cryptography procedure.

*3) Man in the Middle Attack*

One of the most important weaknesses of the RSA algorithm is the possibility of the failure in Man in the Middle attack. In the suggested model the possibility of failure in face with Man in the Middle attack has been reach to 0% because of using an Agent.

The attacker can attacks by being in the middle of Data Owner-Agent or Data Applicant-Agent or Data Owner-Data Applicant. However in none of these cases that attacker can broke the encryption and access to the cloud server because of dual encryption and secure transmission between all entities.

*4) Discrete Logarithm Attack*

In the proposed model, by using AES for the main data the possibility of discrete logarithm attack is decreased. Furthermore, the key of AES algorithm is re-encrypted with RSA-2048 that increases the rate of efficiency in face with this attack.

## VI. CONCLUSION

Regarding to the importance of security issues in cloud computing environments, an efficient and reliable user authentication and data protection model was presented in this paper to increase the rate of reliability in this emerging technology. Accordingly, two encryption procedures were established in an independent middleware (Agent) to perform the process of user authentication, access control, and data protection in cloud servers. AES was used as a symmetric cryptography algorithm in cloud servers and RSA was used as a asymmetric cryptography algorithm in Agent servers. The theoretical evaluation of the proposed model showed that the ability of resistance in face with possible attacked and unpredictable event was enhanced considerably in comparison with similar models because of using dual encryption and an independent middleware during user authentication and data protection procedures.